\documentclass{article}

\usepackage{graphicx}
\usepackage[format=plain, labelfont=sf, textfont=sf]{caption}
\usepackage{color}
\usepackage{enumitem}
\usepackage[colorlinks, linkcolor=black]{hyperref}

\usepackage[style=numeric, maxcitenames=3, abbreviate=true, url=false, isbn=true, backend=biber, url=true]{biblatex}
\usepackage{perpage}
\MakePerPage{footnote}
\bibliography{paper}

\begin{document}

\vspace{0.5cm}

\section*{Electronic Laboratory Notebook: A lazy approach}

\section*{Paper Authors}

1. Schubotz, Simon $^{1,2}$ (schubotz@ipfdd.de);\\
2. Schubotz, Moritz $^{3}$; \\
3. Auernhammer, Günter K. $^{1}$ (auernhammer@ipfdd.de);

\section*{Paper Author Roles and Affiliations}
$^1$ Leibniz-Institut für Polymerforschung Dresden e.V.\unskip, Hohe Stra\ss e 6\unskip, Dresden\unskip, 01069\unskip, Germany\\
$^2 $ Technische Universität Dresden\unskip, Helmholtztra\ss e 10\unskip, Dresden\unskip, 01062\unskip, Germany\\
$^3 $ FIZ Karlsruhe\unskip, Franklinstra\ss e 11\unskip, Berlin\unskip, 10587\unskip, Germany\\

\section*{Abstract}
Good research data management is essential in modern-day lab work. 
Various solutions exist that are either highly specific or need a significant effort to be customized appropriately. 
This paper presents an integrated solution for individuals and small groups of researchers in data-driven deductive research.
Our electronic lab book generates itself out of notes and files, which are generated by one or several experiments.
The generated electronic lab book is then presented on a Django-based website.
The automated gathering of metadata significantly reduces the documentation effort for the lab worker and prevents human error in the repetitive task of manually entering basic meta-data.
The skilled user can quickly adapt the electronic lab book software to his needs because the software employs widely used open-source software libraries with active communities and excellent documentation. 

\section*{Keywords}
Electronic Laboratory Notebook, Data management, Research data management, Scientific software, Automated data processing

\section*{Introduction}

\begin{figure}
	\centering
	\includegraphics[width=1\linewidth]{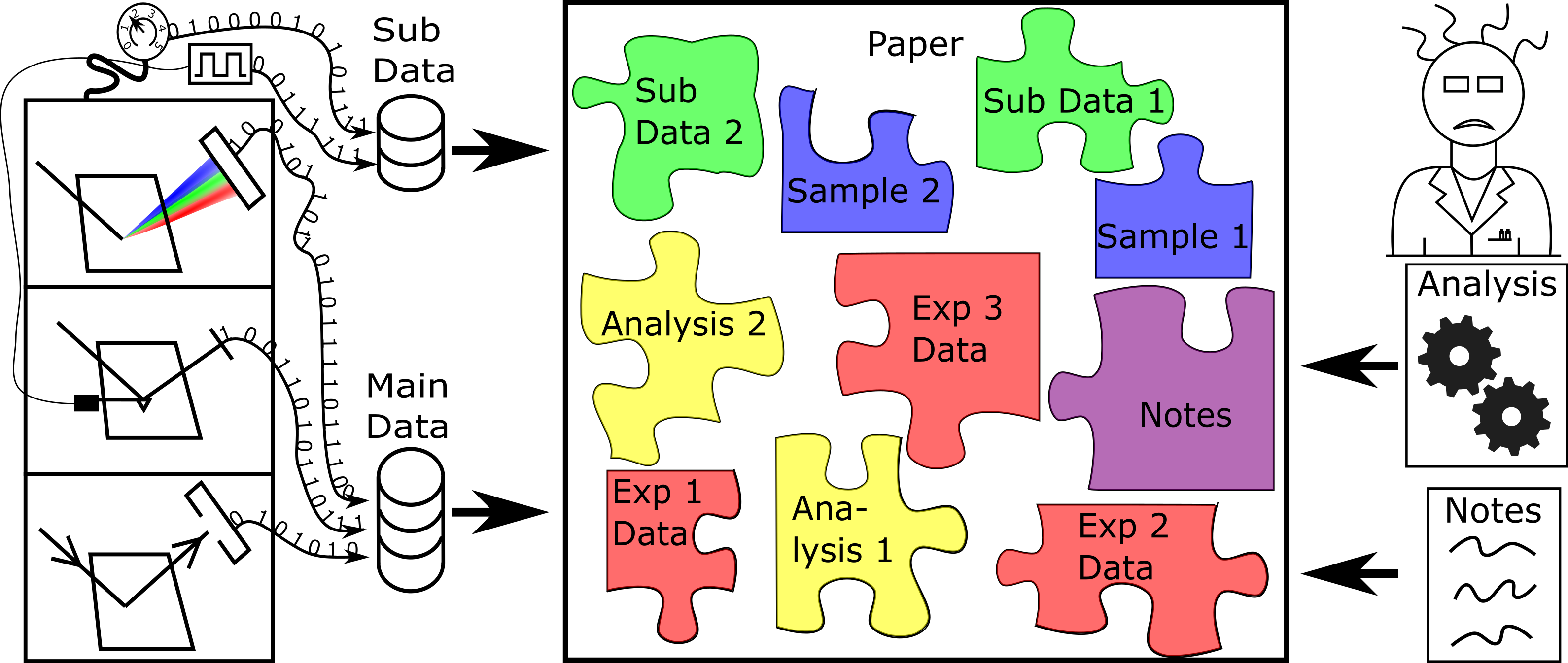}
	\caption{When performing multiple experiments on multiple samples, much data is produced.
	This data has to be structured, analyzed, and potentially linked with some notes or other data.
	In our approach, we try to streamline this process.}
	\label{fig:problem}
\end{figure}

Good research data management should primarily allow the researcher to navigate her or his data easily. 
Additionally, it is an essential prerequisite to archiving and later potential reuse of the data. 
Large-scale experiments like Cern~\cite{Albrecht2019} or KaTriN~\cite{Chilingaryan2010} have significant capabilities in storing and processing data in a streamlined way. 
For small-scale experiments, these capabilities are limited.
Many scientists still rely on handwritten notes.
For the analysis, the notes have to be manually linked to the data.

However, there are approaches for electronic lab books (also often Electronic Laboratory Notebook, short ''lab book'' in this work), but those are often highly specialized~\cite{Bauch2011,CARPi2017,SchultzeMotel2019}.
However, there are approaches for electronic lab books (also often Electronic Laboratory Notebook, short ''lab book'' in this work), but those are often highly specialized.
Another issue is that it is difficult for users to modify lab books because they do not use standard software packages.
Those circumstances make it hard to adapt those lab books to specific needs.
The adaption of the lab book enables a very tight bonding to experiments, reducing the amount of tedious documentation work. 
Particularly for experiments without a built-in data management system, this is useful.
For example, in a self-build experimental setup, data management could be done with the lab book, resulting in a better user experience and possibly reducing the licensing fees by using our open-source lab book.

One of the key differences to other lab books, e.g., those mentioned above, is that we create the lab book directly from the primary data (primarily measurement or simulation data of experiments).
The primary data has to be saved in a particular file structure from which we can already retrieve some metadata.
The automated retrieval of metadata significantly reduces the amount of work, especially if many files have to be treated separately.

Some experiments have multiple data streams.
We call the data from the main device "main data" and data from supporting devices "sub data" (\autoref{fig:problem}).
For example, one could try to measure the optical properties of a sample at different pressures.
In this case, the main data would be the optical data and the sub data the pressure.
The entries of the main- and sub experiments (and parts of the metadata) can then be automatically generated from the data files.
This way, we can handle multiple data streams from the main experiment and form the supporting experiment, which are generated by multiple independent devices.
In addition, it is always possible to manually add or modify data and its metadata.

In our research, we are investigating different samples of polymer brushes with different properties, like different polymer chain length.
By performing experiments on the sample, the properties of the sample, like its wetting properties, can be changed \cite{Schubotz2021}.
For the data interpretation, it is important to keep track of the history of the samples, i.e., which experiments were performed on the sample at which time.
With this lab book, we can filter after one sample and see the samples` history, helping us understand why the sample's properties have been changed.
Currently, our research produces several thousand of these main data entries per year.
To create all these entries in a database manually would require a lot of time.
For a systematic analysis of the experiments, the possibility to filter is essential, e.g., filtering according to experimental parameters, the experiment type, the sample, observations, etc. 

\begin{figure}
	\centering
	\includegraphics[width=1\linewidth]{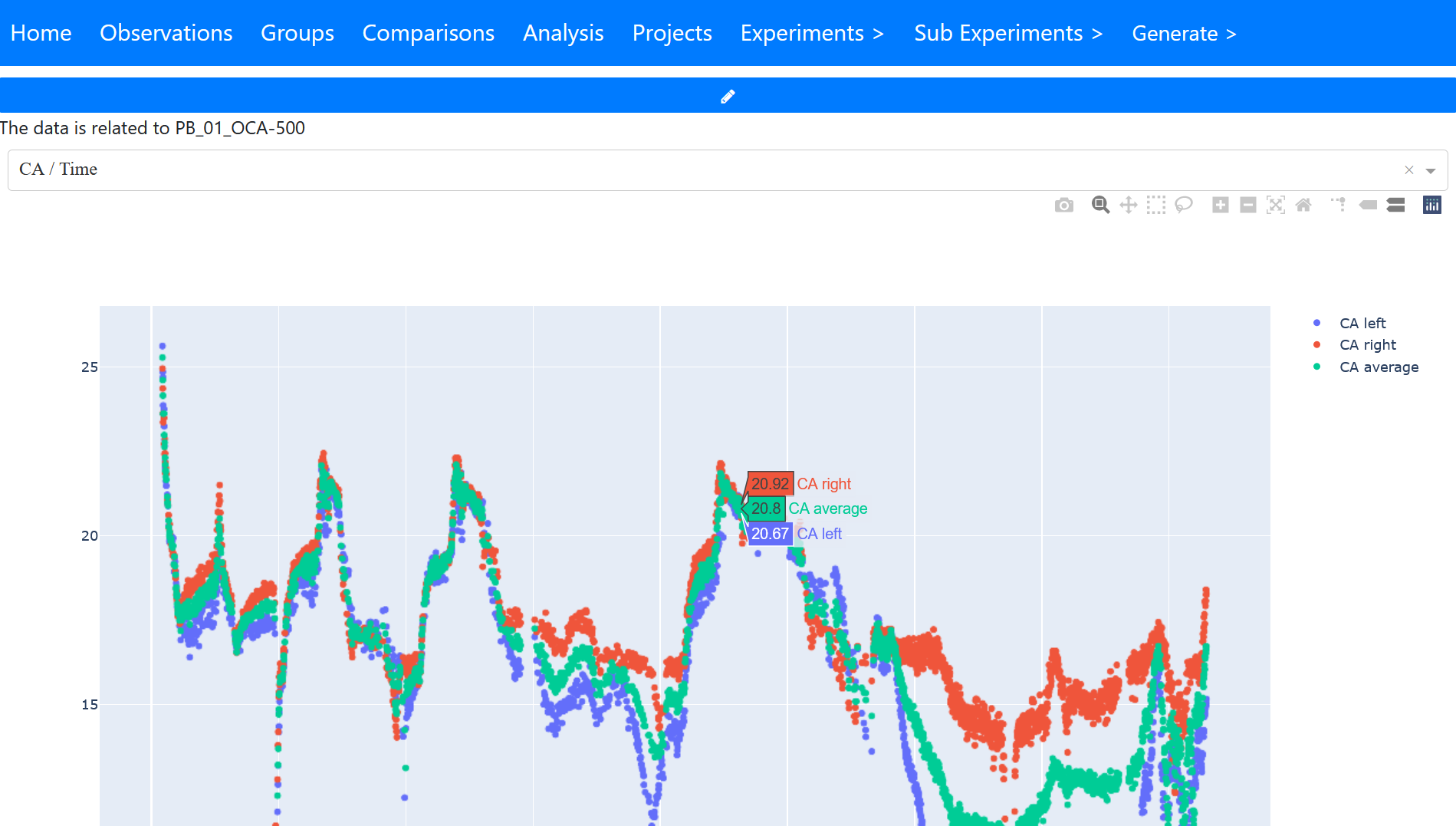}
	\caption{Interactive representation of some experimental data.}
	\label{fig:Image_of_plot}
\end{figure}

From our research, we deduce the following required features of the lab book.
\begin{enumerate}
    \item It should be easy and fast to load hundreds or even thousands of files into the database if they are saved in a structured way.
    \item Metadata should be deduced from the structure the file is saved in and added automatically to the database.
    \item Files that are related should be automatically linked in the database, e.g., main and sub data.
    \item There should be a report about the files that are automatically loaded and linked.
    \item Experiments that produced data, in a regular table format (pandas \footnote{\url{https://pandas.pydata.org/}} compatible) should be shown in form of an interactive plot (\autoref{fig:Image_of_plot}). 
The plot should also be correlated to other data like the sub data.
\end{enumerate}

\section*{Implementation and architecture}
\begin{figure}
	\centering
	\includegraphics[width=1\linewidth]{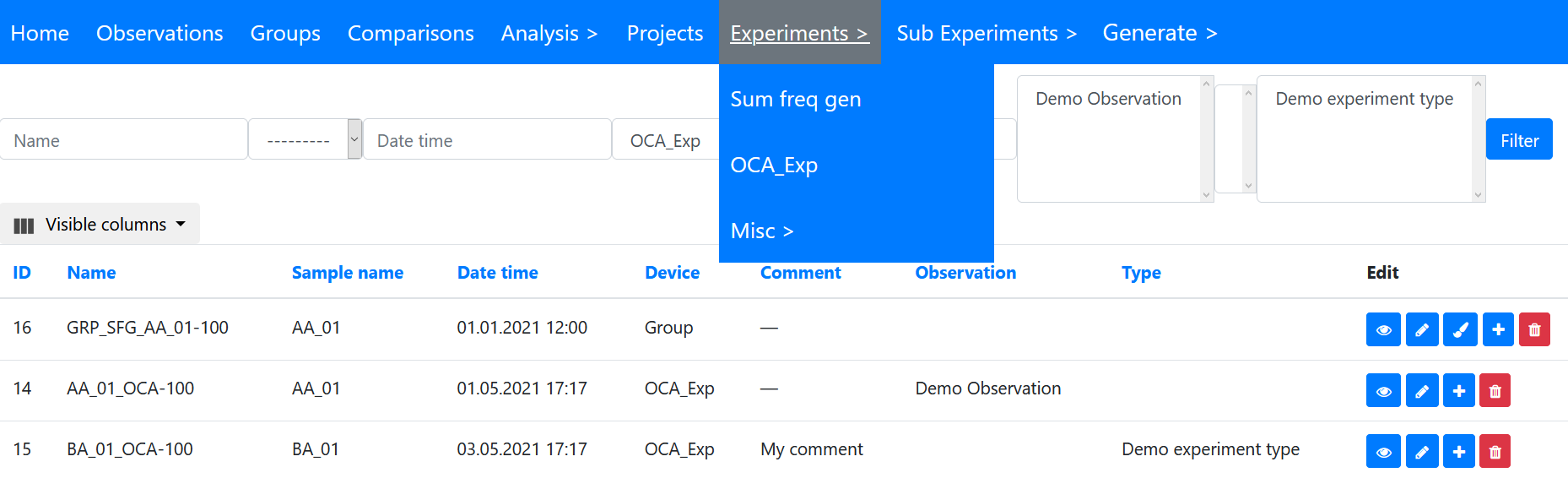}
	\caption{Overview of all experiments that have been performed.}
	\label{fig:Experiments}
\end{figure}

We chose Django as a framework~\cite{Bronger}, because it is written in python, which is the most popular programming language at the moment.
Our Django program consists of 5 Apps: Exp\_Main, Exp\_Sub, Analysis, Lab\_Dash, and Lab\_Misc.

The primary data is linked to the metadata and gets a persistent identifier (ID) when loading data.
Over this ID, the data can be quickly accessed or presented in an interactive graph.
By joining the primary data and the metadata in a database, the data becomes more accessible because one can filter after specific criteria in the metadata and thus keep the overview  \autoref{fig:Experiments}.
The main data can be further enhanced by linking the ID to the sub data ID and their analysis results.

 \begin{figure}
 	\centering
 	\includegraphics[width=1\linewidth]{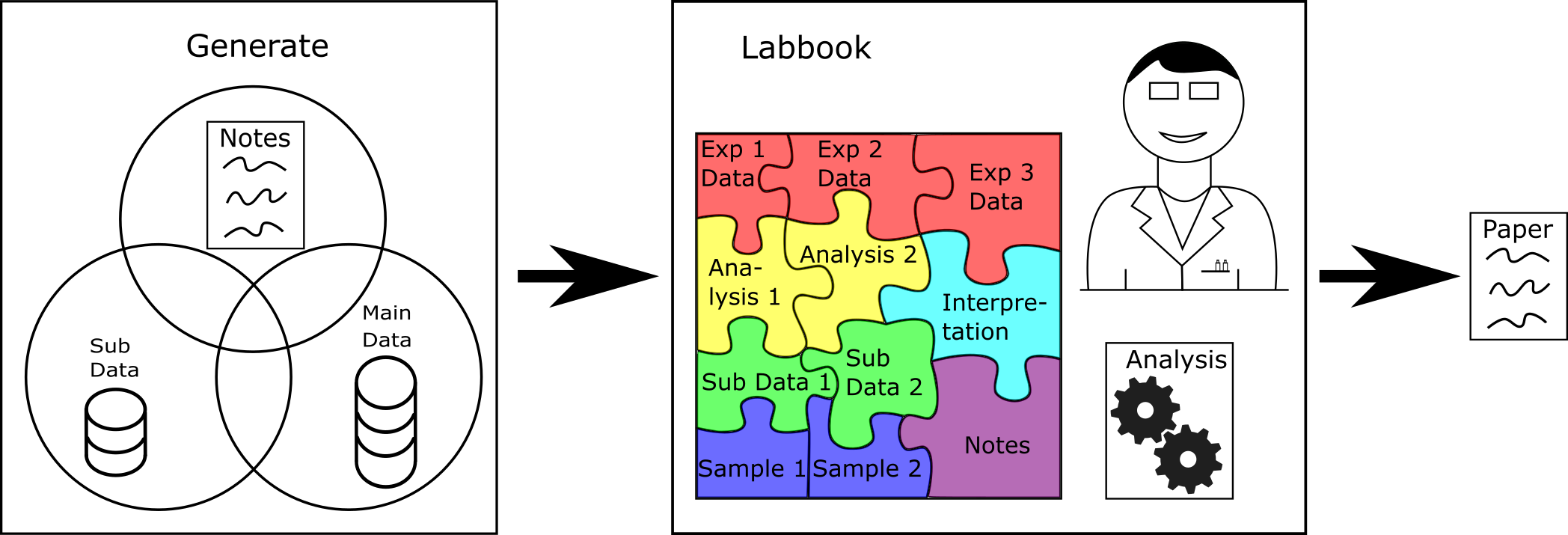}
 	\caption{A significant part of the lab book is generated out of the files of the experiments.
	 Joining the experiments with the notes written into the lab book during the experiments gives a good overview of what was done.
	 After running the data analysis, the user can enter the results and the interpretation into the lab book.
	 Keeping the data, the analysis, and its interpretation linked helps the finding of new scientific results.}
 	\label{fig:Solution}
 \end{figure}

After the experiments the gathered data has to be stored.
For large amounts of data it is useful to organize the data in the filesystem hierarchy.
Building up on this idea a special folder structure is implemented, which contains the date, sample name, and the experiments name (a detailed example can be found in reuse potential section).
By using this special structure, it is possible to automatically generate database entries out of primary data that contains metadata like the date, sample, link to the file, or basic information that is written into the file or filename \autoref{fig:Solution}.
That saves time for the lab worker because the information has only to be entered once and is afterward available in a structured and searchable way.
The metadata retrieved from the folder structure acts as the backbone of the lab book; further metadata can then be added manually or via scripts.

The entries of the experiments are automatically given a persistent ID, which users can use to link these experiments to all kinds of other information.
For example, the sub experiment can be correlated to the main experiment via the timestamp of the data files.
The linkage of these experiments then happens over the ID of the corresponding entries. 
Additionally, other links could be established, e.g., to link to the analysis results, other main experiments, observations, etc.
This is a far more elegant and flexible way to establish, even multiple, structures in the whole data than saving experiments in certain folders, where the folders themself act as the structure.
Relevant information during the experiment can be added by a separate entry and can later be easily correlated to the experiments over the timestamp and sample name.

Django creates the database out of models that have to be defined in Django.
For the entries of the experiments, the model is chosen in the following way.
An experimental base model inherits its properties from all the specific experimental models.
The data relevant for all experiments, like the date, name, and additional metadata, are stored in the base model. 
The experiments' specific data, like wavelength, exposure time, are stored in a specific experiments model.
Thus, the experiment models are always an extension of the experimental base model.
That approach simplifies the complexity of the lab book a lot.

Individual analysis scripts can enhance the primary data.
The primary data and the analysis data can be related via their IDs.
By including the metadata, the analysis can be very specific. 
Results of the analysis can be fed back to the metadata.
Over a jupyter \footnote{\url{https://jupyter.org/}} interface, a script-driven communication with the database and all the Django integrated functions is possible.
This simplifies the development of analysis routines because all web interface features are also available for jupyter scripts, resulting in high flexibility.
The jupyter extension could also extract specific data, e.g., plots or tables, which can be handy for publications.

\begin{figure}
	\centering
	\includegraphics[width=1\linewidth]{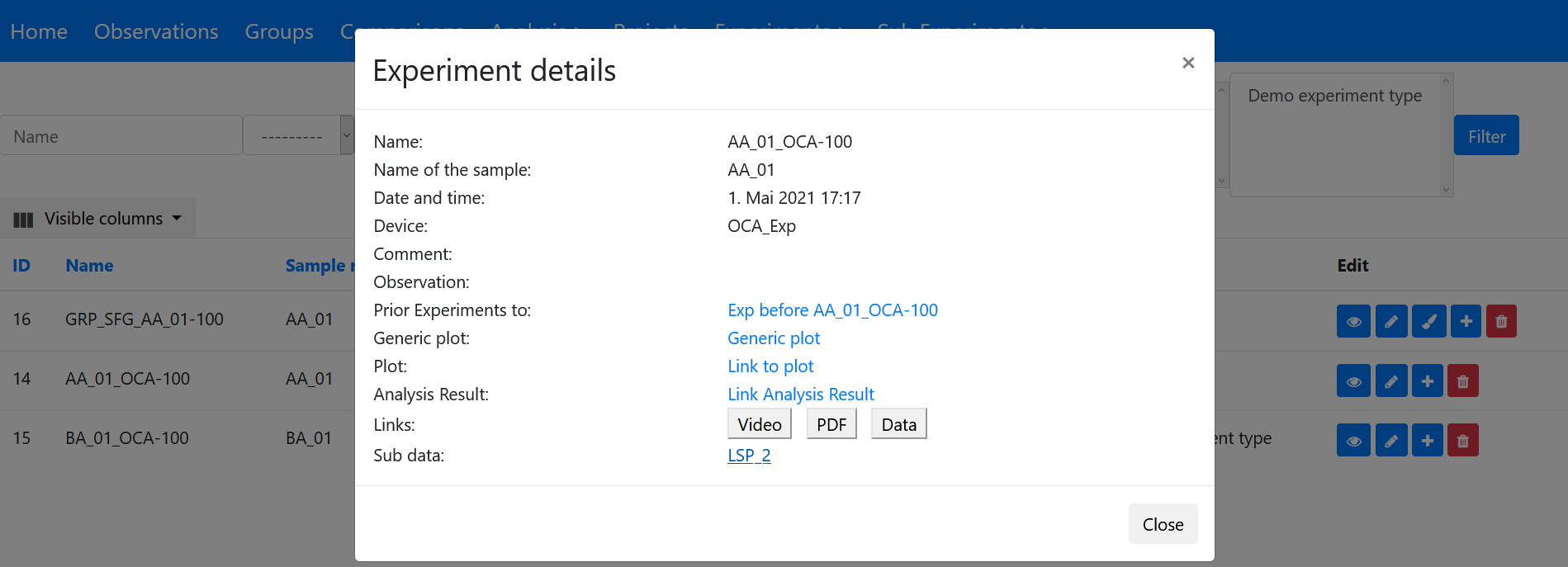}
	\caption{Details of the experiment.}
	\label{fig:Details}
\end{figure}

For a first overview, the data that is stored in the Exp Base model is represented in a table-styled way (\autoref{fig:Experiments}).
In a detailed view, special links related to the experiment can be accessed, for example, the interactive plot, the supporting experiments, and more (\autoref{fig:Details}).
In the table view, the entries can be modified or deleted.
Later, we want to couple the modification rights to the user management system that is integrated into Django.
This would then prevent malicious interference with the entries.

Additionally, users that store their data in (private) GitHub repositorys, can protect their data by a decentralized trusted timestamping (DTT) mechanism.
This happens the OriginStam.org API.
A cryptographic one-way hash of the data is generated and send to the OriginStamp API.
This hash is then combined with other hashes and integrated into the BitCoin Blockchain and thus immutably stored.
With this hash, we can prove that we had the data when the hash was published, which happens once a day.
Only publishing the hash rather than the entire data set prevents others from making use of the data before the measurement is finalized, properly peer-reviewed, and adequately described.
This method is incredibly viable for users unwilling or unable, e.g., due to legal restrictions, to publish their entire data set.
GitHub has an option to require signed commits, and OriginStamp can seamlessly be integrated via a Webhook with almost no effort \cite{Hepp2018,Wortner2019}.
The signed commits options ensures that the authorship of the data can be claimed with a public-private key method.

For users that can not use GitHubs` webkook functionality (or the equivalent fuinctionaltiy in GitLabs), we recommend to combine the user authentication with the DTT and integrate the hash generation, signing and backup functionality into the lab book.
However, this is out of scope for us.

\section*{Quality control}
For the quality control, we used the Django testing framework.
Several unit tests are performed, and also an integration test.
On the tested machines, there were no error messages during testing.

\section*{(2) Availability}
\vspace{0.5cm}

\section*{Operating system}
Windows 10

Runs with docker, thus also on unix systems.
In windows the functionality is higher but the installation effort is also larger.

\section*{Programming language}
Python 3.8.2

\section*{Dependencies}
django, django-tables2, django-bootstrap3, django-filter, django-tables2-column-shifter, django-bootstrap-modal-forms, django-widget-tweaks, django\_plotly\_dash, channels, bootstrap4, dash, dash-renderer, dash-html-components, dash-core-components, plotly, numpy, pandas, daphne, redis django-redis, channels-redis, django-mptt, xlrd, dbfread, kaleido, django-datatables-view, jupyter, ipython, django-extensions and scipy

\section*{Software location:}

{\bf Archive}
\begin{description}[noitemsep,topsep=0pt]
	\item[Name:]  Electronic-Laboratory-Notebook
	\item[Persistent identifier:] \href{https://archive.softwareheritage.org/swh:1:dir:c6b704517578897e3a7feba2489fcd6c9548bf7a}{swh:1:dir:c6b704517578897e3a7feba2489fcd6c9548bf7a}
	\item[Licence:] Apache License 2.0
	\item[Publisher:]  Softwareheritage
	\item[Date published:] 18/08/2021
\end{description}

{\bf Code repository}
\begin{description}[noitemsep,topsep=0pt]
	\item[Name:] GitHub
	\item[Persistent identifier:] \href{https://github.com/ag-gipp/Electronic-Laboratory-Notebook}{ag-gipp/Electronic-Laboratory-Notebook}
	\item[Licence:] Apache License 2.0
	\item[Date published:] 18/08/2021
\end{description}

\section*{Language}
English

\section*{(3) Reuse potential}
The program can be used by anyone who wants to access their data in a more structured way.
Because it is open-source, the customizability is extensive. 
One could even introduce more users and lift it to a server and thus run collaborative projects.
However, this would require some work of the user who is familiar with Django. 
Here, we will introduce the first steps for single-user customization.

\subsection*{Add new file to exsisting model and sample}
If we want to add a new file to the lab book, we can copy it to the folder structure.
One example file is already present in this commit.
The file is saved at  \nolinkurl{01\_Data/01\_Main\_Exp/01\_OCA\_35\_XL/20210201/Probe\_BA\_01/171700\_osz\_wasser\_laengest.png}
From this path, the program extracts the information that the measurement was recorded on the first of February at 17:17:00, and the sample BA\_01 was used in the experiment OCA.
Due to performance optimizations, the file will be ignored if it is older than five days or has the wrong file extension.
The user configures the allowed file extensions.
To add the example, change the date in the folder structure to yesterday by renaming the according to folder.
Navigate in the header to ''Generate'' and click on ''Main'' and then on ''Generate entries''.
When going back to the experiments, there should be a new entry with different data.
By clicking on the eye and then on Video, the data is displayed as an image.
If you have files that do not have the time written in front of them, consider using the give\_file\_times module in Exp\_Main/Generate.py
We decided not always to generate the time automatically to allow manually link the data using the exact times.

\subsection*{Add a new sample}
To add a new sample, navigate to \url{http://127.0.0.1:8000/admin/} and log in with the username: admin and password: admin
Go to Lab\_Misc and click on samplebrushpnipaamsi and then add.
It is essential to give each sample a proper name consisting of two letters and two numbers separated by an underscore, e.g., CC\_01.
Then click save.
Thereafter, you can use this name in the folder structure and add a sample as described above.
It is probably helpful to define your own model for your sample, which inherits Sample\_Blanks.
For further details about models, visit \url{https://docs.djangoproject.com/en/3.1/topics/db/models/}

\subsection*{Add a new experiment}
To add a new experiment, navigate to Exp\_Main/models.py and add your model, which inherits ExpBase.
For a start, you can follow the example of the OCA model.
The model name should consist of 3 capital letters.
After the Django migration, the model should be visible in the admin.
To tell the program where the files of this model are stored, navigate in the admin to Exp path and create a new entry with the path file ending and the abbreviation, which consists of the same three capital letters as in the model.

If you use a Windows system, it can be beneficial to run Django without docker, since only with Windows is it possible to open local files directly from the web browser.
If files can not be accessed directly by the web browser, users of Unix systems need to copy the link into the local file browser manually.
The rest works normally.

\subsection*{Support}
If support is needed do not hesitate to use the appropriate tools of GitHub.

\section*{Funding statement}
Simon Schubotz thanks the Deutsche Forschungsgemeinschaft (DFG) for funding of project 422852551 (AU321/10-1) within the priority program 2171.
Günter K. Auernhammer was funded by the Deutsche Forschungsgemeinschaft (DFG, German Research Foundation)–Project-ID No. 265191195–SFB 1194-A02. 

\section*{Competing interests}
The authors declare that they have no competing interests.

\printbibliography

\end{document}